\documentclass[twocolumn]{new-aiaa} 
\usepackage[utf8]{inputenc}
\usepackage{todonotes}

\usepackage{graphicx}
\usepackage{amsmath}
\usepackage{subfigure}
\usepackage{longtable,tabularx}
\usepackage{lineno}
\usepackage{svg}
\usepackage{printlen}

\title{Development of transitional Reynolds number correlation and assessment of RANS for predictions of bypass transition}

\author{Carlos A. Gonzalez \footnote{Ph.D. candidate, Dept. of Mechanical Engineering, Stanford University.}, Rahul Agrawal \footnote{Ph.D. candidate, Dept. of Mechanical Engineering, Stanford University. AIAA Student Member.}}\affil{Center for Turbulence Research, Stanford University, Stanford, CA 94305, USA}
\author{Xiaohua Wu \footnote{Professor of Mechanical and Aerospace Engineering, Royal Military College of Canada, Canada. AIAA Associate Fellow. } }
\affil{Royal Military College of Canada, Kingston, \\ Ontario K7K 7B4, Canada}

\begin{document}
\maketitle

\begin{abstract}
We present direct numerical simulations (DNSs) of bypass transition over a flat plate with inlet freestream turbulence intensity levels of $0.75\%$, $1.5\%$,  $2.25\%$,  $3.0\%$ and $6.0\%$, respectively. A new definition of the transition intermittency is proposed based on the mean skin friction. Based on these, we develop an intermittency correlation to predict flow transition. The proposed model is consistent with the classical correlation of Abu-Ghannam and Shaw and reasonably predicts transition Reynolds number (within 10.8\% error) for the experiments of \citet{fransson2020effect}. Accompanying Reynolds-averaged Navier-Stokes (RANS) simulations for our DNS cases simulations are performed. The RANS results are sensitive to the specification of the inlet turbulence length scale and overpredict (underpredict) the growth of the integral flow scales across the boundary layer during transitional stages when the inlet freestream turbulence is low (high), respectively. \\ 

\end{abstract}

\section{Introduction}
Engineers have strong and sustained interests in predicting the location of boundary-layer transition in the flows over aircraft wings, jet-engine turbofan airfoils, compressors, and turbine blades. Transition modeling has remained an active research field over the past half-century. Representative approaches for engineering transition prediction include data correlations, Reynolds-averaged Navier-Stokes (RANS) turbulence models, a RANS turbulence model coupled with an intermittency equation, a stand-alone intermittency equation, and laminar fluctuation energy (for additional details, see \citet{dick2017transition} and \citet{durbin2018some}). \\

\begin{figure}[!ht]
    \centering    \includegraphics[width=0.9\linewidth]{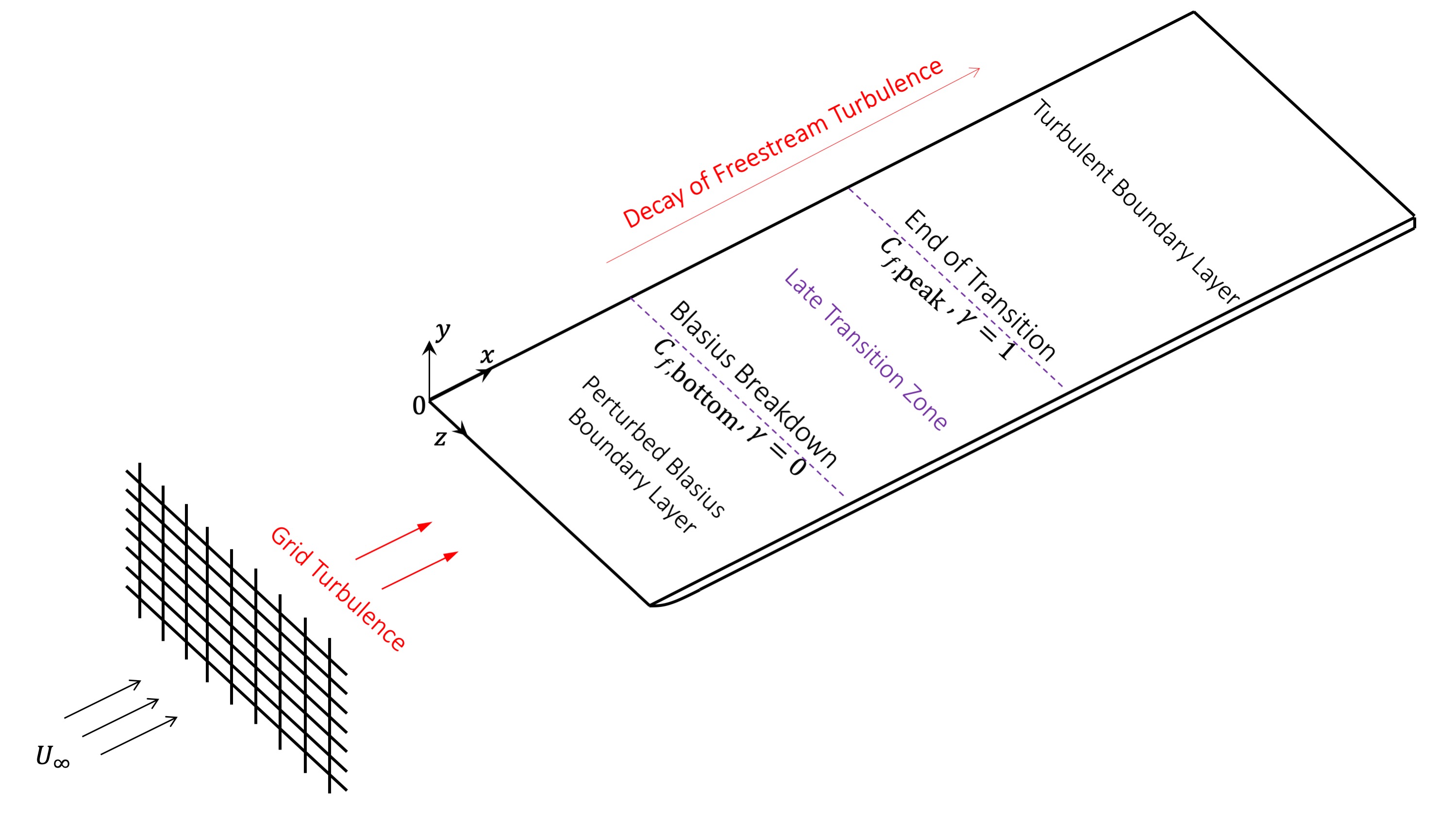}
    \caption{Sketch of boundary-layer bypass transition in the narrow sense.}
    \label{fig:sketch_01}
\end{figure}

\noindent
An essential step in transition prediction is the evaluation of the predictive ability of a given RANS/correlation model in a zero-pressure-gradient, smooth-wall, flat-plate boundary layer under isotropic freestream turbulence intensity at the inlet ($\mathrm{FSTI}_{in}$) levels between $0.5\%$ and $6\%$. Such flows were referred to as boundary-layer bypass transition in the narrow sense (also see Figure~\ref{fig:sketch_01}) in \citet{wu2017transitional} to distinguish them from bypass transition arising from other sources, such as separation bubble, roughness element, or very high-level freestream turbulence.

\noindent
Often, the experiment of \citet{roach1990influence} is used for benchmarking and evaluating transition prediction models. \citet{roach1990influence} reported wind-tunnel experiments in three conditions: case T3A, with upstream turbulence intensity $3.5\%$; case T3B, with upstream turbulence intensity $6.5\%$; and case T3A-, with upstream turbulence intensity $0.8\%$. A salient feature of their experiments is that their skin-friction coefficient $C_f$ data collapse onto the Blasius solution before laminar-layer breakdown. Several transition models developed since then have used Roach's skin-friction coefficient $C_f$ data for benchmarking and tuning (for instance, see \citet{westin1997application}, \citet{suzen2000modeling}, \citet{menter2006transition}, \citet{durbin2012intermittency}, \citet{ge2014bypass}, \citet{menter2015one}). \\  

\noindent
Despite its popularity, the dataset of \citet{roach1990influence} lacks documentation of the development history of the FSTI length scales over the boundary layer. Further, there are only a few measured data points during the transition between $C_f$ departing from the laminar distribution and attaining the turbulent level, which may be undesirable from a modeling perspective, especially given the significant differences between FSTI of different experiments. This indicates a potential need to generate a more thorough database for transition model benchmarking and calibration. \\

\noindent
To this end, several direct numerical simulation (DNS) databases exist. However, the skin friction in some of these existing DNSs departs from the Blasius solution profile earlier than desired in the early transitional region. The present study aims to improve this limitation. To the authors' knowledge, this database represents a comprehensive range of $\mathrm{FSTI}_{in}$ simulated under one computational framework. The rest of this note is organized as follows. In Sections \ref{sec:method} and \ref{sec:cfwudns}, we present our DNS database, the skin friction of which agrees well with the Blasius solution until the transition region. Section \ref{sec:improved} demonstrates the equivalence between the conventional definition of the intermittency factor (based on temporal signals, see \citet{fransson2020effect}) and a new definition proposed herein based purely on the mean skin friction. We also present a new transition prediction correlation over flat plate flows involving the intermittency factor, Reynolds number, and FSTI scales. Section \ref{sec:rans} assesses RANS predictions of the transition for our DNSs and highlights some key sensitivities of RANS predictions to the choice of turbulent length scales. Finally, concluding remarks are offered in Section \ref{sec:conclude}.    
 
\section{Method}
\label{sec:method}
Here, we summarize the computational approach to perform five DNSs of boundary-layer bypass transition over a flat plate in the narrow sense.
Inlet FSTI levels ($\mathrm{FSTI}_{in}$) for these flows correspond to $0.75\%$ (denoted as WM075), $1.5\%$ (denoted as WM150), $2.25\%$ (denoted as WM225), $3.0\%$ (denoted as WM300) and $6.0\%$ (denoted as WM600). For brevity, the reader is referred to \citet{wu2017transitional} for a discussion of the domain size, inflow turbulence 
($\Vec{v}_{\text{isotropic}}$) generation, boundary conditions, and numerical discretization of our flow solver. \\

\noindent
Let $x, \, y, \, z$ denote the flow's streamwise, wall-normal, and spanwise directions, respectively. Free-stream turbulence is introduced at the inlet over the wall-normal range $15\theta_{\text{in}} <y<L_{\text{y,iso}}$, ($\theta_{in}$ is the constant inlet boundary layer momentum thickness), above which is only a uniform flow $U_{\infty}$ (without fluctuations), and below which is only the Blasius velocity profile (without fluctuations). At the streamwise exit, domain height $L_{y}$ is equivalent to $24.31$, $9.05$, $7.36$, $7.24$, and $6.66$ local boundary-layer thickness ($\delta_{\text{exit}}$) for cases WM075, WM150, WM225, WM300 and WM600, respectively. 
Similarly, $L_{\text{y,iso}}$ is equivalent to 9.28 and 2.54 local boundary-layer thickness $\delta_{\text{exit}}$ in WM075 and WM600, respectively. \\

\noindent
At the top of the computational domain, we apply the Dirichlet boundary condition to the wall-normal velocity ($v=v_{\text{Blasius}}$) and zero vorticity condition to the streamwise and spanwise velocities ($\partial u/\partial y=\partial v/\partial x$, $\partial w/\partial y=\partial v/\partial z$). Given the substantial distance between the top surface and the wall, we hypothesize that the effect of the top boundary condition on boundary-layer development should be minimal. \\

\begin{figure}
    \centering
    \includegraphics[width=1.0\linewidth]{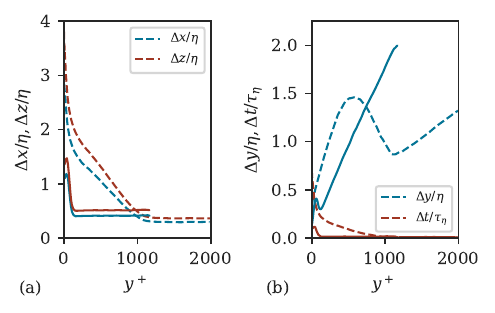}
    \caption{DNS spatial and temporal resolutions measured by local Kolmogorov scales in WM150. Dashed lines correspond to the streamwise station $Re_{\theta} = 2000$ and solid lines to the streamwise station $Re_{\theta} = 500$.}
    \label{fig:dns_resolution_wm150}
\end{figure}

\noindent
Figure~\ref{fig:dns_resolution_wm150} shows the streamwise and spanwise grid resolutions normalized by the local Kolmogorov length scale ($\eta=(\nu^{3}/\varepsilon)^{1/4}$ where $\nu$ is the kinematic eddy viscosity of the fluid, and $\varepsilon$ is the local rate of viscous dissipation of turbulence kinetic energy evaluated from the DNS) for WM150 flow. Wall-normal distributions of $\Delta x/\eta$ and $\Delta z/\eta$ at two representative streamwise stations are shown: $Re_{\theta}=500$ during the transition and $Re_{\theta}=2000$ in the fully turbulent region. It is seen from Figure~\ref{fig:dns_resolution_wm150}(a) that during the transition, the horizontal plane grid resolutions are less than $2\eta$, and after the transition, they are less than $4\eta$. The near-flat profiles at larger values of $y^{+}$ indicate the freestream region outside the boundary layer. Figure~\ref{fig:dns_resolution_wm150}(b) shows the normalized wall-normal resolution ($\Delta y/\eta$) and the temporal resolution ($\Delta t/\tau_{\eta}$) in WM150, where $\tau_{\eta}=(\nu/\varepsilon)^{1/2}$ is the local Kolmogorov timescale. The temporal resolution is less than $\tau_{\eta}$, thus likely resolving the small-scale temporal dynamics of the flow. The wall-normal resolution is finer than $2\eta$ within the boundary layer and the freestream, highlighting a well-resolved computational domain.  
 
\section{Skin-friction profiles from DNS}
\label{sec:cfwudns}

Here, we present the processed skin friction data from our DNSs. For the sake of brevity, other quantities such as boundary-layer thickness ($\delta$), displacement thickness ($\delta^{*}$), shape factor ($H$), wall-pressure fluctuation ($p_{w,rms}^{'}$), wall-shear stress fluctuation ($\tau_{w,rms}^{'}$) are not presented. However, the streamwise growth of these data with the streamwise ($Re_x$),  momentum thickness ($Re_\theta$) or the friction-based ($Re_\tau$) based Reynolds numbers are accessible from the \href{https://ctr.stanford.edu/about-center-turbulence-research/research-data}{Center for Turbulence Research website}. Additionally, the database contains the wall-normal variations of the mean velocity ($\overline{u}$) and of turbulent stresses ($u_{rms}^{'}$, $v_{rms}^{'}$, $w_{rms}^{'}$, $\overline{u^{'}v^{'}}$, $p_{rms}^{'}$, where overbar indicates averaging),
and total shear ($\tau$) in outer units ($y/\delta$) or in viscous units ($y^{+}$) at selected streamwise stations covering the early, late and post-transition stages. The statistics were sampled on the fly at every time step over two convective flow-through times across the domain. The reader is referred to \citet{wu2017transitional,gonzalez2024benchmarks} for a discussion on the accuracy of the collected statistics. \\

\begin{figure}[!ht]
    \centering
    \includegraphics[width=1\linewidth]{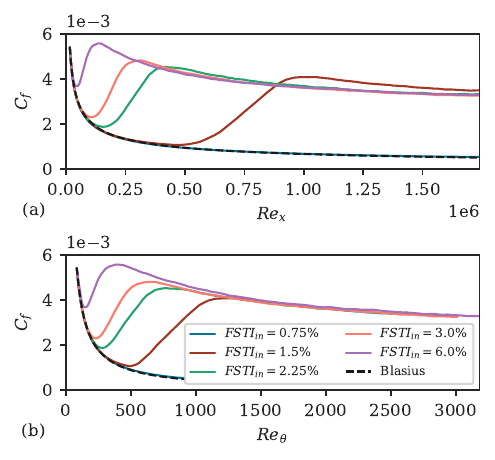}
    \caption{Skin-friction coefficient as a function of (a) streamwise Reynolds number and (b) momentum-thickness Reynolds number in the five boundary-layer DNS cases.}
    \label{fig:cf_Rex}
\end{figure}

Figure~\ref{fig:cf_Rex}(a) presents the development of the skin friction, $C_{f}$ with the streamwise Reynolds number, $Re_{x}$. Beginning at the inlet, all cases agree well with the Blasius skin-friction prediction over an extended streamwise Reynolds number. Departure from the Blasius solution is generally observed only shortly upstream of the minimum $C_{f}$ location. At the lowest $\mathrm{FSTI}$ (WM075), the flow does not transition by the exit of the chosen computational domain. \\

The development of $C_{f}$ with $Re_{\theta}$ is shown in Figure~\ref{fig:cf_Rex}(b). Shortly after reaching a peak plateau, the WM225 $C_{f}$ profile collapses onto that of the WM300, suggesting the downstream WM225 boundary layer is turbulent with a small lingering transitional effect. Similarly, shortly after reaching its peak plateau, the WM150 profile collapses onto those of WM225 and WM300, implying the downstream WM150 boundary layer is turbulent with a small transitional effect (the peak plateau of skin friction is generally chosen to indicate the completion of transition). Even for the WM600 flow, the $C_f$ prediction is generally similar to the three immediately lower $\mathrm{FSTI}$ cases beyond the peak $C_f$ plateau. However, the WM600 profile does not exactly collapse onto the other profiles, potentially suggesting that the increased $\mathrm{FSTI}$, especially beyond 6\% may introduce non-negligible disturbances to the viscous sublayer of the turbulent boundary layer.

\section{Improved intermittency correlation and comparison with literature}
\label{sec:improved}

In this section, we propose a new definition of the intermittency function ($\gamma$) (which is used as a marker for the completion of flow transition) based on the mean skin friction data. We highlight that a typical measure of $\gamma$ is the temporal definition based on the probability that the flow at some location and time is turbulent \cite{pope2021turbulentflows}. The proposed $\gamma$ definition only includes the measurement of the mean skin friction instead of a temporal record of the turbulent velocity fluctuations. Further, we will utilize the proposed definition of $\gamma$ in developing a transition Reynolds number correlation in this section.    

\subsection{Skin-friction based measure of intermittency}
\label{sec:gammadef}

\begin{figure}
    \centering
    \includegraphics[width=1.0\linewidth]{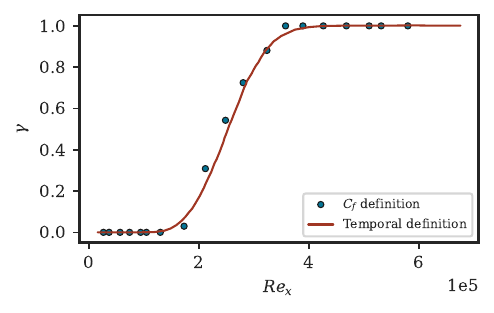}
    \caption{Intermittency versus $Re_x$. The solid line denotes the typical temporal definition of intermittency, while the symbols denote computing intermittency using Equation \ref{eq:gamma-from-cf}.}
    \label{fig:gamma-cf-vs-temporal}
\end{figure}

We propose to use the following definition based on the mean skin friction, 
\begin{equation}
    \gamma = \frac{C_f - \max(C_f)}{\max(C_f) - \min(C_f)}.
    \label{eq:gamma-from-cf}
\end{equation}
where $\max(C_f)$ is the value of the skin friction at the peak plateau, which is a marker of the completion of the transition. Using the DNS data of \citet{mamidala2022comparative} and the corresponding experiments of \citet{fransson2020effect}, we show in Figure~\ref{fig:gamma-cf-vs-temporal} that the skin-friction-based definition of intermittency agrees well with the conventional, temporally computed definition.

\subsection{Correlation for transition Reynolds number }
\begin{figure}
    \centering
    \includegraphics[width=0.7\linewidth]{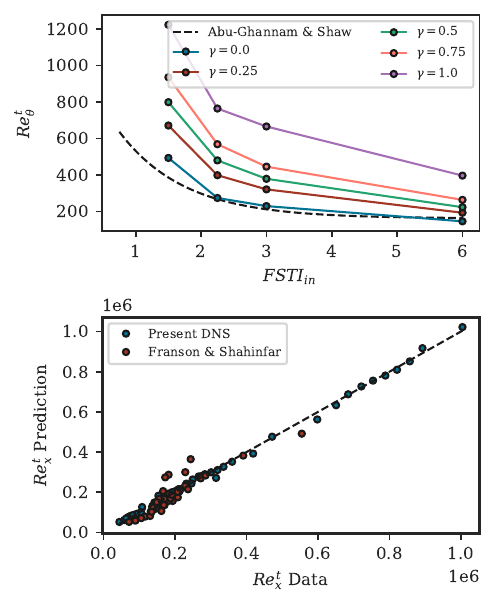}
    \caption{(a) Transition location measured in $Re^t_{\theta}$ plotted against the inlet freestream turbulence intensity level. Here, we report transition locations for $\gamma = 0.25 n$ for $0 \leq n \leq 4$. (b) Comparing transition prediction results from Equation \ref{transition-correlation} to measured results from present simulations and the experimental data set of \citet{fransson2020effect}. The experimental data set has a variety of conditions for $\mathrm{FSTI}_{in}$ and free stream turbulence integral length scales. The transition threshold is chosen to be $\gamma = 0.5$.  }
    \label{fig:agsComp_and_correlation}
\end{figure}

Building upon the DNS database, we propose a new correlation for predicting boundary-layer transition over flat plate flows. Phenomenologically, we invoke a model form, 

\begin{equation}
    Re^t_x, \, Re^t_{\theta} = f(\mathrm{FSTI}_{in}, \frac{L_{in}}{\delta_{in}}, \gamma),
\end{equation}
where $Re^t_x, \, Re^t_{\theta}$ denote the freestream and momentum thickness Reynolds numbers at the transition location.  $L_{in}/\delta_{in}$ is the inlet freestream turbulence integral length scale ($\sim k^{3/2}/\epsilon$ where $k$ is the turbulent kinetic energy) normalized by the inlet boundary-layer thickness, and $\gamma$ is the desired intermittency factor (based on skin-friction definition) to demarcate as the transition location, respectively. \citet{fransson2020effect} have previously shown that transition location is sensitive to the length scales associated with freestream turbulence. In developing our new transition correlation, we observed that the classical correlation of \citet{abu1980natural}

\begin{equation}
    Re^t_{\theta} = 163 + \exp(6.91 - \mathrm{FSTI}_{in}),
    \label{ags-eq}
\end{equation}
reasonably predicts the location where $\gamma = 0$ in our DNS data across the various freestream turbulence intensities (see Figure~\ref{fig:agsComp_and_correlation}(a)). Based on this agreement, we propose a model functional form to ensure that the predicted transition $Re_\theta$ collapses to a form similar to \citet{abu1980natural} relation when $\gamma = 0$. This is enforced by invoking the form 
\begin{equation}
    Re^t_{x} = P_1\left(\frac{L_{in}}{\delta_{in}}\right) \cdot \exp(P_2(\gamma)) \cdot g(\mathrm{FSTI}_{in}),
    \label{transition-correlation}
\end{equation}
where $P_1$ and $P_2$ are polynomial functions and $g$ is a function similar to Equation \ref{ags-eq}. It is highlighted that the proposed model is fitted for the transition value of $Re_x$, principally because of the ease of its definition. Moreover, this equation is valid only for $\gamma \in [0, 1]$, which are the range of sensible values for the intermittency function. Curve fitting tests (only on the DNS data) demonstrated that the fitting cubic polynomials in Equation \ref{transition-correlation} for $P_1$ and $P_2$ were sufficient to predict the transition. The exact forms of $P_1$, $P_2$ and $g$ are 
\begin{equation}
\begin{split}
    P_1\left(\frac{L}{\delta}\right) = -2.41\cdot 10^{-4} \left(\frac{L}{\delta}\right)^3 + \\ 2.77\cdot10^{-3}\left(\frac{L}{\delta}\right)^2 + 5.19\left(\frac{L}{\delta}\right) + 271.6,
\end{split}
\label{eqn:p1}
\end{equation}

\begin{equation}
    P_2(\gamma) = 1.68 \gamma^3 - 2.84 \gamma^2 + 1.92 \gamma + 2.72,
\end{equation}
and
\begin{equation}
    g(\mathrm{FSTI_{in}}) = 5.23 - \exp(5.92 - 1.05 \; \mathrm{FSTI_{in}}).
\end{equation}

As a further validation outside the ``training set'',  the transition location predictions from our model for the experiments of \citet{fransson2020effect} are presented in Figure~\ref{fig:agsComp_and_correlation}(b). A reasonable agreement is observed, both for the DNS and the experimental data with the average error in the prediction of $Re^t_x \approx 10.8\%$. 

\noindent

\section{RANS predictions of the DNS cases}
\label{sec:rans}

Here, we present the accompanying RANS calculations for the DNS cases. These simulations were computed using OpenFOAM's simpleFoam solver. For RANS simulations, we chose the size of the computational domain to be identical to the DNS. A laminar Blasius velocity profile was prescribed at the inlet to match the DNS. Although not shown, grid convergence studies were performed for RANS, and convergence was achieved when the near-wall, wall-normal resolution, $y^+ < 1$ for all cases. The results presented herein correspond to the grid converged cases. \\

\begin{figure}
    \centering
    \includegraphics[width=1.0\linewidth]{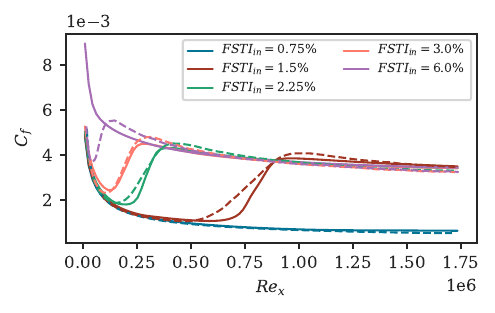}
    \caption{Solid lines depict skin friction versus $Re_x$ for RANS results, and dashed lines depict the corresponding DNS results. The length scale used in these results' boundary condition for $\omega$ is $\mathcal{L} = \delta_{in}$.}
    \label{rans-d99-cf}
\end{figure}

\noindent
The $k-\omega$ shear-stress transport (SST) turbulence model \citep{menter1994two} and $\gamma-Re_\theta$  \citep{langtry2005transition} transition model are used. Here $\omega$ refers to the specific rate of dissipation of turbulent kinetic energy. For brevity, we do not pursue a detailed investigation of the sensitivity of the predictions to different RANS models. These models require inlet values for $k$, $\omega$, $\gamma$, $Re_{\theta}$, and the RANS eddy viscosity, $\nu_t$.
A discussion of the boundary conditions for these RANS models can be found in \citet{langtry2009correlation}. The equation for the boundary condition for $\omega$ is given by Equation \ref{omega-bc} where $C_\mu = 0.09$. This equation leaves a free parameter, a reference length scale $\mathcal{L}$, associated with the flow turbulence.
\begin{equation}
    \omega = \frac{\sqrt{k}}{C_\mu^{0.25} \mathcal{L}}
    \label{omega-bc}
\end{equation}
We highlight that implementation differences of the $k-\omega$ SST model can lead to different recommended values of $\mathcal{L}$. For example, in wall-bounded flows, ANSYS recommends setting $\mathcal{L} = 0.4\delta_{in}$ \citep{Ansys} where $\delta_{in}$ is the boundary layer thickness at the inlet. On the contrary, COMSOL recommends $\mathcal{L} = 0.09\delta_{in}$ \citep{comsol}. Although not shown, overall we found that in OpenFOAM the value of this parameter that best reproduces the DNS skin-friction profiles for our cases is $\mathcal{L} = \delta_{in}$. The skin-friction curves resulting from a choice of $\mathcal{L} = \delta_{in}$ are shown in Figure~\ref{rans-d99-cf}.  For this length scale, $\mathcal{L}$, the WM600 case is fully turbulent throughout the whole computational domain and does not predict the onset of transition\footnote{Although not shown, our numerical experiments suggest that tuning $\mathcal{L}$ can lead to the prediction of the onset of transition at the 6\% $\mathrm{FSTI_{in}}$ level but the same value of $\mathcal{L}$ does not produce correct $C_f$ curves for the lower $\mathrm{FSTI}_{in}$ cases.}. \\

\begin{figure}
    \centering
    \includegraphics[width=1.0\linewidth]{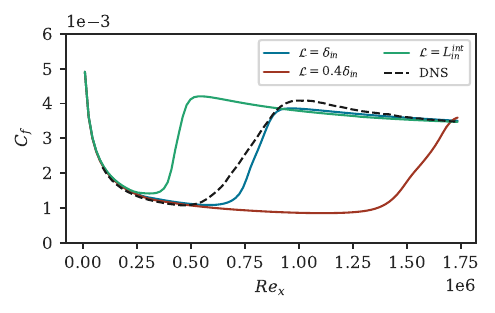}
    \caption{Skin-friction versus $Re_x$ results for RANS computations at $\mathrm{FSTI}_{in} = 1.5\%$. The large sensitivity in the transition location is tied to the choice of $\mathcal{L}$ used to set the inlet boundary condition for $\omega$.}
    \label{fig:rans-L-sensitivity}
\end{figure}

\begin{figure}
    \centering
    \includegraphics[width=1.0\linewidth]{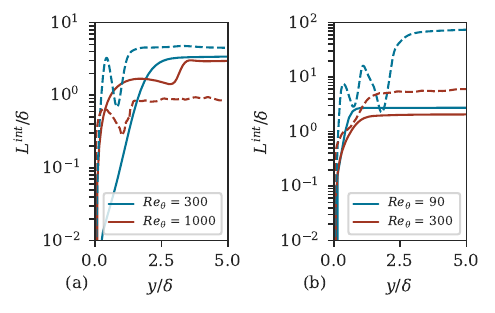}
    \caption{Normalized integral length scale versus normalized wall-normal height at two streamwise stations for (a) $\mathrm{FSTI}_{in}=1.5\%$ and (b) $\mathrm{FSTI}_{in}=6\%$. Dashed lines correspond to DNS results, and solid lines correspond to RANS. The length scale used for $\omega$ for these simulations is $\mathcal{L} = \delta_{in}$.}
    \label{fig:rans-integral-length-scale}
\end{figure}

\noindent
In Figure~\ref{fig:rans-L-sensitivity}, we demonstrate the sensitivity of the predictions from the RANS model to the choice of $\mathcal{L}$ for $\mathrm{FSTI}_{in} = 1.5\%$ case. We use $\mathcal{L} = \delta_{in}$, $\mathcal{L} = 0.4\delta_{in}$, and $\mathcal{L} = L^{int}_{in}$ where $L^{int}$ is the integral length scale of the inlet freestream turbulence. It is apparent that the location of transition and the magnitude of the skin-friction rise are affected by the choice of $\mathcal{L}$ which highlights a potential shortcoming of the chosen RANS transition models.  Further, in Figure~\ref{fig:rans-integral-length-scale}, we find that performing RANS with $\mathcal{L}=\delta_{in}$ leads to an underprediction of the wall-normal variation of the integral length scale of the boundary layer at the pre-transition streamwise station, corresponding to $Re_\theta=\{300, \, 90\}$ for $\mathrm{FSTI}_{in} = \{1.5\%, \, 6\% \}$ respectively. For the lower $\mathrm{FSTI}_{in}$, at a representative ``during transition'' station ($Re_\theta \sim 1000$), the integral scales are overpredicted compared to DNS. On the contrary, for the higher FSTI flow at the ``during transition'' station ($Re_\theta \sim 300$), the RANS still underpredicts the growth of the integral scales. These are consistent with the observation (in Figure \ref{rans-d99-cf}) that RANS predicts a delayed and early transition at the lowest and highest $\mathrm{FSTI}_{in}$ respectively. Further, these are also consistent with the proposed $Re^t_x$ correlation in that for $L/\delta \sim \mathcal{O}(1)$, an over (under) prediction in $L/\delta$ results in a delayed (early) transition, $Re^t_x$, prediction respectively. \\

\begin{figure}
    \centering
    \includegraphics[width=1.0\linewidth]{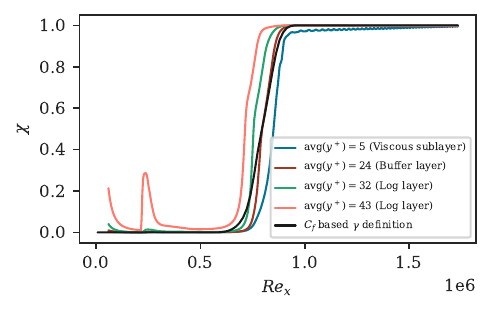}
    \caption{Intermittency versus $Re_x$ from RANS of the $\mathrm{FSTI}_{in} = 1.5\%$ flow. Colored lines denote $\gamma$ from the intermittency transport equation at fixed wall-normal distances. Legend entries correspond to their respective mean $y^+$ value. The black line denotes the RANS $C_f$ based intermittency.}
    \label{fig:gamma-cf-vs-transport}
\end{figure}

For comparing the consistency of the transported values of the intermittency factor ($\gamma$) from RANS with the $C_f$ (from RANS) based definition (proposed in Section \ref{sec:gammadef}), we plot $\chi$ defined as

\begin{equation}
    \chi = \frac{\gamma - \max(\gamma)}{\max(\gamma) - \min(\gamma)},
    \label{eq:chi-eqn}
\end{equation}
at different wall-normal locations for $\mathrm{FSTI}_{in}=1.5\%$ flow.\footnote{This re-normalization of $\gamma$ to $\chi$ aids the comparison of $\gamma$ across different wall-normal heights. Further, we also exclude the first 15 (near the inlet) values (out of 500) of $\gamma$ in this computation since these values are numerically polluted.} As apparent in Figure~\ref{fig:gamma-cf-vs-transport}, the $\chi$ values below the log layer are delayed compared to the $C_f$ based definition. In the early part of the log layer, we see the best agreement with the $C_f$ based definition, but nevertheless, the slope and shape of the curve are in disagreement. Finally, in the highest plotted portion of the log layer, we observe that $\chi$ rises early, with an erroneous spike at $Re_x\approx0.23\cdot10^6$\footnote{Although not shown, $\gamma$ at even higher wall-normal locations exhibit similar erroneous spikes.}. These characteristics are undesirable given the observations in Section \ref{sec:gammadef}. 

\section{Conclusions}
\label{sec:conclude}

In this note, we present a newly simulated set of comprehensive DNS of bypass transition with varying levels of inlet freestream turbulence intensity. Using this database, a new transition correlation has been developed to predict the onset of flow transition based on a user-specified threshold of the intermittency function ($\gamma$). We show that our correlation is similar to the existing correlation of \citet{abu1980natural} for $\gamma=0$. We further include the information of freestream turbulence length scales and desired intermittency threshold to predict flow transition. Beyond the computed DNS, we present reasonable predictions (average error of $10.8\%$) in the prediction of the streamwise, transition Reynolds number ($Re^t_x$) for the experiments of \citet{fransson2020effect}. Further, we demonstrate a similarity between the conventional, temporal probability-based definition of the intermittency function ($\gamma$) and the proposed mean skin friction-based definition. \\

\noindent
From a modeling perspective, accompanying RANS simulations were conducted using OpenFOAM. These calculations demonstrate that the predictions of transition and skin-friction for the chosen models ($k-\omega$ turbulence model and the $\gamma-Re_\theta$ transition model) are sensitive to the specified length scale at the domain inlet. It is shown that the RANS calculations overpredict (underpredict) the growth of the integral length scales of the turbulence across the boundary layer at a representative ``during transition'' station when the inlet freestream turbulence intensity is at the lowest, $\approx 1.5\%$ (highest, $\approx 6 \%$) respectively. Finally, the transported intermittency factor in RANS predicts a transition onset relative to the $C_f$ definition that is dependent on the height at which $\gamma$ is probed (in the viscous or the buffer layer, the transition is delayed, but an earlier transition in the logarithmic layer). 

\section*{Acknowledgements}

This work was pursued during the 2024 Center for Turbulence Research Summer Program. C.A.G. and R.A. were supported by the NASA Transformational Tools and Technologies Program (grant \#80NSSC20M0201). R.A. also acknowledges support from Boeing Research and Technology (grant \#2024-UI-PA-100) and the Franklin P. and Caroline M. Johnson Fellowship at Stanford University. X.W. is supported by NSERC Discovery Grant and Digital Alliance Canada. The authors gratefully acknowledge helpful discussions with Prof. Parviz Moin, Dr. Bumseok Lee, and Dr. Tomek Jaroslawski. 


 \end{document}